\documentclass[sigplan,nonacm,screen]{acmart}
\usepackage[export]{adjustbox}
\usepackage{listings}
\usepackage{xspace}
\usepackage{graphicx}
\usepackage{glossaries}
\usepackage{subcaption}
\usepackage{multirow}
\usepackage{booktabs}
\definecolor{LightGray}{gray}{0.9}

\definecolor{SafeGreen}{HTML}{1b9e77}
\definecolor{SafeRed}{HTML}{d95f02}
\definecolor{SafeBlue}{HTML}{7570b3}

\glsdisablehyper

\def \txone {ThunderX-1\xspace}

\newacronym{acpi}{ACPI}{Advanced Configuration and Power Interface}
\newacronym{afu}{AFU}{Accelerator Function Unit}
\newacronym{asic}{ASIC}{Application-Specific Integrated Circuit}
\newacronym{atf}{ATF}{ARM Trusted Firmware}
\newacronym{bdk}{BDK}{Board Development Kit}
\newacronym{bist}{BIST}{built-in self-test}
\newacronym{bmc}{BMC}{Baseboard Management Controller}
\newacronym{bram}{BRAM}{Block RAM}
\newacronym{capi}{CAPI}{Coherent Accelerator Processor Interface}
\newacronym{capp}{CAPP}{Coherent Accelerator Processor Proxy}
\newacronym{ccip}{CCI-P}{Core Cache Interface}
\newacronym{ccix}{CCIX}{Cache Coherent Interconnect for Accelerators}
\newacronym{cpld}{CPLD}{Complex Programmable Logic Device}
\newacronym{cxl}{CXL}{Compute eXpress Link}
\newacronym{dac}{DAC}{Digital Analog Converter}
\newacronym{dma}{DMA}{Direct Memory Access}
\newacronym{eci}{ECI}{Enzian Coherence Interface}
\newacronym{ept}{EPT}{Extended Page Table}
\newacronym{fiu}{FUI}{FPGA Interface Unit}
\newacronym{fmc}{FMC}{FPGA Mezzanine Card}
\newacronym{fpga}{FPGA}{Field Programmable Gate Array}
\newacronym{gpu}{GPU}{Graphics Processing Unit}
\newacronym{gsync}{\textsc{GSync}}{Global Synchronization}
\newacronym{hbm}{HBM}{High-Bandwidth Memory}
\newacronym{hpc}{HPC}{High-Performance Computing}
\newacronym{i2c}{I\textsuperscript{2}C}{Inter-Integrated Circuit}
\newacronym{ic}{IC}{Integrated Circuit}
\newacronym{ipc}{IPC}{Inter-Process Communication}
\newacronym{ipi}{IPI}{Inter-Processor Interrupt}
\newacronym{ipmi}{IPMI}{Intelligent Platform Management Interface}
\newacronym{llc}{LLC}{Last-Level Cache}
\newacronym{mmio}{MMIO}{Memory-Mapped I/O}
\newacronym{mmu}{MMU}{memory management unit}
\newacronym{mpsoc}{MPSoC}{Multiprocessor System-on-a-Chip}
\newacronym{mpx}{MPX}{Memory Protection Extensions}
\newacronym{msi}{MSI}{Message-Signalled Interrupt}
\newacronym{nic}{NIC}{network interface adaptor}
\newacronym{nvme}{NVMe}{NVM Express}
\newacronym{ocapi}{OpenCAPI}{Open Coherent Accelerator Processor Interface}
\newacronym{pae}{PAE}{Physical Address Extensions}
\newacronym{pcb}{PCB}{Printed Circuit Board}
\newacronym{pcie}{PCIe}{PCI Express}
\newacronym{pio}{PIO}{Programmed I/O}
\newacronym{pmbus}{PMBus}{Power Management Bus}
\newacronym{psci}{PSCI}{Power State Coordination Interface}
\newacronym{psl}{PSL}{POWER Service Layer}
\newacronym[longplural=page table entries]{pte}{PTE}{page table entry}
\newacronym{qpi}{QPI}{QuickPath Interconnect}
\newacronym{rdma}{RDMA}{Remote Direct Memory Access}
\newacronym{rpc}{RPC}{Remote Procedure Call}
\newacronym{rtc}{RTC}{Real Time Clock}
\newacronym{sata}{SATA}{Serial ATA}
\newacronym{sgx}{SGX}{Software Guard Extensions}
\newacronym{smbus}{SMBus}{System Management Bus}
\newacronym{smm}{SMM}{System Management Mode}
\newacronym{smmu}{SMMU}{System Memory Management Unit}
\newacronym{smc}{SMC}{Secure Monitor Call}
\newacronym{soc}{SoC}{System-on-Chip}
\newacronym{som}{SoM}{System-on-Module}
\newacronym{spl}{SPL}{System Protocol Layer}
\newacronym{tap}{TAP}{Test Access Port}
\newacronym{tdp}{TDP}{Thermal Design Power}
\newacronym{tfa}{TF-A}{Trusted Firmware-A}
\newacronym{tlb}{TLB}{Translation Lookaside Buffer}
\newacronym{tpu}{TPU}{Tensor Processing Unit}
\newacronym{ttbr}{TTBR}{Translation Table Base Register}
\newacronym{uart}{UART}{universal asynchronous receiver-transmitter}
\newacronym{uefi}{UEFI}{Unified Extensible Firmware Interface}
\newacronym{upi}{UPI}{Universal Path Interconnect}
\newacronym{vfpga}{vFPGA}{Virtual FPGA}
\newacronym{vpu}{VPU}{Video Processing Unit}
\newacronym{xmpu}{XMPU}{Xilinx Memory Protection Unit}
\newacronym{xppu}{XPPU}{Xilinx Peripheral Protection Unit}

\AtBeginDocument{%
	\providecommand\BibTeX{{%
			\normalfont B\kern-0.5em{\scshape i\kern-0.25em b}\kern-0.8em\TeX}}}

\newcommand{\Modified}{\textsc{Modified}\xspace}

\newcommand{\Exclusive}{\textsc{Exclusive}\xspace}
\newcommand{\Shared}{\textsc{Shared}\xspace}
\newcommand{\Invalid}{\textsc{Invalid}\xspace}

\newcommand{\micros}{$\mu$s\xspace}

\begin{document}

	\title{Rethinking Programmed I/O for Fast Devices, Cheap Cores, and Coherent Interconnects}

        \author{Anastasiia Ruzhanskaia}
        \affiliation{%
            \institution{Systems Group, D-INFK, ETH Z{\"u}rich}
            \city{Z{\"u}rich}
            \country{Switzerland}
        }

        \author{Pengcheng Xu}
        \affiliation{%
            \institution{Systems Group, D-INFK, ETH Z{\"u}rich}
            \city{Z{\"u}rich}
            \country{Switzerland}
        }

        \author{David Cock}
        \affiliation{%
            \institution{Systems Group, D-INFK, ETH Z{\"u}rich}
            \city{Z{\"u}rich}
            \country{Switzerland}
        }

        \author{Timothy Roscoe}
        \affiliation{%
            \institution{Systems Group, D-INFK, ETH Z{\"u}rich}
            \city{Z{\"u}rich}
            \country{Switzerland}
        }

	\newcommand{\fast}{2F2F}

	\begin{abstract}

Conventional wisdom holds that an efficient interface between an OS
running on a CPU and a high-bandwidth I/O device should use \gls{dma}
to offload data transfer, descriptor rings for buffering and queuing,
and interrupts for asynchrony between cores and device.

In this paper we question this wisdom in the light of two trends:
modern and emerging \emph{cache-coherent interconnects} like CXL3.0,
and workloads, particularly \emph{microservices} and \emph{serverless
computing}.  Like some others before us, we argue that the assumptions
of the DMA-based model are obsolete, and in many use-cases
\emph{programmed I/O}, where the CPU explicitly transfers data and
control information to and from a device via loads and stores,
delivers a more efficient system.

However, we push this idea much further.  We show, in a real hardware
implementation, the gains in latency for
fine-grained communication achievable using an open cache-coherence
protocol which exposes cache transitions to a smart device, and
that throughput is competitive with \gls{dma} over modern
interconnects. We also demonstrate three 
use-cases: fine-grained RPC-style invocation 
of functions on an accelerator, offloading of operators in a streaming
dataflow engine, and a network interface targeting serverless
functions, comparing our use of coherence with both traditional
\gls{dma}-style interaction and a highly-optimized implementation
using memory-mapped programmed I/O over PCIe.
	
\end{abstract}

	\maketitle

	\pagestyle{plain}

	\section{Introduction}\label{sec:introduction}

Modern interfaces between CPUs and high-performance devices like
\glspl{nic}, \glspl{gpu}, etc. are designed to optimize
\emph{throughput for large transfers}.  Built over a
\gls{pcie} interconnect, \emph{descriptor rings} in memory hold
queues of requests and completions, and the device uses
\gls{dma} to read and write both descriptors and payload
data.  This design, along with the underlying \gls{pcie} interconnect,
trades off latency for small transactions in favor of high throughput
for large ones.

We question this consensus in the light of
workloads which depend for performance not on bulk throughput, but the
cumulative latency of small transactions between CPU and device:
closely-coupled accelerators for irregular workloads, or
\gls{rpc} using small messages.  We do this in the context
of emerging \emph{cache-coherent} peripheral interconnects like
CXL.mem 3.0~\cite{cxl3cache}.

The continuing rise in \gls{pcie} bandwidth means, for a
given transfer size, the significance of \gls{pcie} \emph{latency}
(time to first byte) is now more of a concern, leading to many
proposals for reducing the overhead of descriptor management (which we
survey in~\autoref{sec:related}).

We propose a more radical approach, aimed at use-cases where
low latency is more important than high maximum throughput:
using \gls{pio} to issue loads and stores to \gls{mmio} registers directly, avoiding
\gls{dma}, descriptors, queues, and interrupts entirely.

Using a real hardware platform which implements a coherent
interconnect between a server-class CPU and a large \glstext{fpga}, we
investigate the trade-offs between \gls{dma} with descriptors,
\gls{pio} directly over a \gls{pcie} interconnect, and \gls{pio} over
a full cache-coherence protocol.

We show that, for small ($\le$ 1KiB) transfers, \gls{pio}
posted \emph{writes} over \gls{pcie} significantly outperform
\gls{dma}, although \emph{reads} are much less efficient since, unlike
writes, they cannot be pipelined and so incur \gls{pcie}'s significant
round-trip latency, in line with other recent critiques of
\gls{dma}~\cite{scenic-route,enso,cc-nic}.

We go further, however, and show how a conventional MESI-like
coherence protocol can be \emph{exploited in novel ways} by
intelligent devices interacting with an unmodified CPU using coherence
messages.  By \emph{relaxing traditional coherence protocol
assumptions} (for example, the independence of cache lines), we can
achieve dramatically more efficient communication between a CPU and a
device. A coherence-based message protocol which avoids \gls{pcie}
significantly outperforms both \gls{dma} and \gls{pio} over \gls{pcie}
and \emph{completely eliminates tail latency}, while delivering the
same or better throughput.

We first provide background and motivation, including
the origins of descriptor-based
\gls{dma}, and why it is time to question these underlying
assumptions.  In \autoref{sec:platform} we describe and
motivate the experimental hardware platform we use for evaluation,
calibrating performance against conventional PC
servers.  We also decribe \gls{pio} with \gls{pcie} here.

In \autoref{sec:opportunities}, we discuss why true cache-coherent
device interconnects like CXL3.0 are fundamentally different from
today's \gls{pcie}, and present a family of protocols that pass
messages with low latency between software on an unmodified CPU, and a
smart device with message-level access to the coherence protocol.
We then present the implementation of these protocols on the hardware platform.

\autoref{sec:evaluation} compares traditional \gls{dma}-based
I/O with \gls{pio} over \gls{pcie} and our new coherence-based
protocols using three use-cases: (1) lightweight, synchronous
local invocation of functionality on a computational accelerator, (2)
I/O to and from an intelligent network interface, and (3) hardware
offload of stream processor operators.
We survey related work in \autoref{sec:related} and conclude with
\autoref{sec:conclusion}.

	\section{Background and Motivation}\label{sec:background}

Our work in this paper is motivated by the interaction between modern
trends in platform interconnect and also the changing ways in which
these interconnects are used, particularly in data center and cloud
servers.

\paragraph{Interconnects and devices:}
In early machines, the CPU interacted with peripherals exclusively
using \emph{\glsfirst{pio}}: the device exposed \gls{mmio}
registers, and the processor issued loads and stores to directly to
these registers to perform both data transfer and control of the
device.

This model made sense when devices were extremely simple, particularly
combined with device interrupts to obviate the need for
the CPU to poll status registers on the device waiting for an I/O
operation to complete.  However, it still requires 
synchronous rendezvous between CPU and device, and moreover
requires all data to be transferred via CPU registers (since the CPU
is copying data between device registers and main memory).

\gls{dma} removed this limitation by allowing the device itself
or a dedicated \gls{dma} controller to copy data while the
CPU continues to execute other code, providing parallelism and partly
decoupling CPU and device.  Further decoupling is provided by
\emph{descriptor rings}: producer/consumer buffers of I/O requests in
memory, read and written by both CPU and \gls{dma}-capable
device.  In this case, interrupts need only be used when the queue of
descriptors becomes full or empty.

Almost every high-speed device today uses this technique,
with minor variations such as the format of descriptor queues,
where they are stored, and whether the queue head and tail are
identified by registers or flags in the descriptors.
This model is optimized for \emph{throughput} -- it is most efficient
when a large volume of data must be transferred.  On modern 
\gls{pcie} interconnects, this consensus results in impressive
bandwidth for applications which transfer data in large chunks, up to
64GiB/s for \gls{pcie} 5.0 x16.

Surprisingly,
increasing \gls{pcie} throughput
over time has not been accompanied by a reduction in
\emph{latency}, which has remained a roughly 1\micros for an
interconnect round-trip message exchange~\cite{pcieswitch}.
For the large transfers used in GPU-derived programming models
(e.g., machine learning workloads), this latency penalty is
insignificant when amortized over the transfer.  Indeed, much
software has adapted to this model imposed by the
hardware~\cite{hardware_lottery}.

However, recent trends have led to proposed \emph{cache
coherent} peripheral interconnects, and the extension of 
processor cache coherence to heterogeneous devices: accelerators,
networking interfaces, and
storage~\cite{OpenCAPI,ccix,cxl3cache,nvlink,chi}.  The more 
far-reaching are fully \emph{symmetric}: devices
are full participants in a distributed directory-based cache
coherence protocol, as in CXL.mem 3.0~\cite{cxl3cache}.

\sloppy
\paragraph{Fine-grained workloads:}
While throughput-oriented workloads involving data transfer between
devices and cores (e.g., AI training and inference) are hugely
important today, we step back in this paper and consider other
workloads that might be a poor fit today for this model, because they
involve fine-grained, frequent interaction between the CPU and another
device. Many irregular workloads have this
property~\cite{gioiosa_pushing_2017,bean_improving_2016,kashimata_cascaded_2019,zeng_graphact_2020}.

Moreover, data center messaging (essentially, \glspl{rpc})
exchange many small messages~\cite{google,dagger}.  While end-to-end
latency is mostly network propagation time in this case, latency
incurred in the end system is a good proxy for key resources (such as
CPU cycles) wasted before invoking the application code, and in
marshaling and sending any reply.

All these workloads incur a significant latency penalty when using \gls{dma}
for data transfer.  Moreover, the assumptions which led to the common
\gls{dma} descriptor-based model do not hold for these workloads.
For example, CPU execution and I/O were decoupled using descriptor
rings because CPU cycles were precious (there being few cores), and
the CPU had plenty of different tasks to work on at any time.

This is no longer true.  Modern servers now have 100s of
cores~\cite{ampereone} and when running data center applications they
are often dedicated to a single application, sometimes with
miscellaneous functions moved to a small
subset~\cite{google-snap}. Most of the cores have nothing to do 
except handle small messages coming from the network.

As a second example: for throughput-oriented workloads \gls{dma} has evolved
to efficiently transfer data to and from main memory without polluting
the CPU cache.  However, for small, fine-grained interactions, it is
important that almost all the data gets into the right CPU cache as
quickly as possible.

All this led us to rethink the \gls{dma}-based model, and whether directly
involving the CPU in data transfers to and from devices might be a
better approach for these scenarios on modern hardware.  We are not
the first to suggest this direction for some
cases~\cite{larsen_reevaluation_2015,chaudhari_evaluating_2017,scenic-route,cc-nic},
but we push the question much further: given a future cache-coherent
interconnect, what \emph{new ways of using \gls{pio}} might
be compelling for data center applications?

We are not proposing abandoning the \gls{dma} model, or
some of its enhancements we survey in~\autoref{sec:related}, for
large-transfer throughput-oriented applications, which can keep riding the current hardware bandwidth curve.  Instead, we argue it
is time to examine complementary alternatives.

	\begin{figure}[t]
	\includegraphics[scale=1]{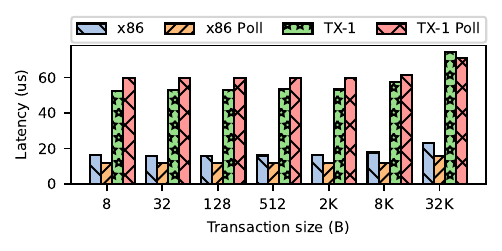}
	\caption{PCIe XDMA invocation latency comparison.}
        \label{fig:ev:together_xdma}
\end{figure}
\begin{figure}[t]
	\includegraphics[scale=1]{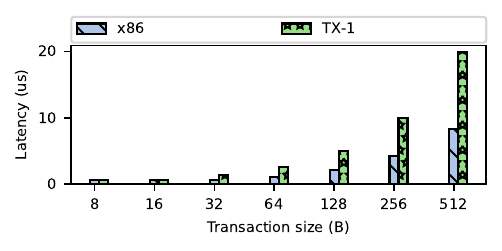}
	\caption{PCIe PIO invocation latency comparison.}
        \label{fig:ev:together_pci}
\end{figure}
\section{Experimental platform}\label{sec:platform}

We investigate the modern trade-offs between \gls{pio}, \gls{dma}, and
cache-coherent interconnects using a real hardware platform rather
than a simulator.  Simulation would be appropriate for detailed
quantitative comparisons of known techniques, but in this paper we
explore unconventional uses of cache coherence and so want to
establish not only that our techniques are performant, but that they
can practically be implemented in a real system.

The Enzian research computer~\cite{Cock:Enzian:2022} can be thought of
as a two-socket NUMA server, where one socket houses a Marvel Cavium
\txone CPU, with a Xilinx XCVU9P \gls{fpga} in the other. The CPU is a
48-core ARMv8 processor running at 2.0GHz, and 128GiB of 2133MT/s DDR4
memory spread across 4 memory controllers.  Each core has a 32KiB,
32-way associative, write-through, physically indexed, physically
tagged, L1 data cache. These are connected to a 16MiB, write-back,
16-way associative, shared L2 last-level cache.  Hardware keeps the L1
caches coherent with the L2, and so in ARM architecture parlance the
point of coherence is the L1 cache, whereas the point of unification
is the L2 cache write-through.   The cache-line size is 128 bytes --
note that this is double the conventional 64-byte line size.

The CPU and \gls{fpga} are connected by the CPU's native inter-socket cache %
coherent protocol, the \gls{eci}, which Enzian implements on the \gls{fpga}.
It is a MOESI-like directory protocol with the physical address
space statically partitioned between the NUMA nodes (CPU and
\gls{fpga}).  It uses 24 bidirectional lanes (organized into 2 12-lane
links), each running at about 10Gb/s, for a theoretical inter-socket
bandwidth of about 30GiB/s.

Both nodes have \gls{pcie} interfaces: the \txone has a \gls{pcie} Gen.3
x8 interface, while the \gls{fpga} has a \gls{pcie} Gen.3 x16 interface.  In
this work we connect these two interfaces together with a cable; this replicates a \gls{pcie} accelerator card and
allows us to compare \gls{pcie} with \gls{eci} in our experiments.

While Enzian provides a real hardware platform with server-class
performance, direct comparison with modern server platforms is not
straightforward: the latter do not offer full cache coherence
between CPU and devices, but have more recent, faster
cores, memory systems, and \gls{pcie}.

We therefore calibrate both
\gls{pio} and \gls{dma} over \gls{pcie} with benchmark
comparisons of Enzian and a modern PC (Intel Core i7-7700 3.6GHz Kaby
Lake, PCIe Gen.3 x16) with 64-byte cache lines connected to an
AMD Virtex UltraScale+ VCU118 card, using the ``write then
read'' experiment detailed in \autoref{sec:eval:invocation}.  The VCU118 uses
the same \gls{fpga} as Enzian, albeit a slightly slower speed gauge.

\paragraph{\gls{dma} performance over \gls{pcie}:}

The first experiment uses the Xilinx XDMA IP and its
descriptor-based protocol to transfer data between CPU and
\gls{fpga} over \gls{pcie}.  \autoref{fig:ev:together_xdma} shows results
for various data transfer sizes, running the Xilinx driver in both
interrupt-driven and polled modes.

XDMA over \gls{pcie} is about 3 times faster on the PC
than on Enzian, while the difference between interrupt-driven and
polling performance is much less significant.  Single transaction
latency is almost constant on both platforms up to the \gls{pcie}
transaction size limit of 4KiB, and then increases.

The performance difference between the machines is due to
several factors.  For small transfers, the CPU overhead of
descriptor setup dominates, and an x86 core is simply faster
than a \txone core.  In addition, the
memory system is somewhat faster on the PC.  The factor of 2
difference in \emph{bandwidth} between the two platforms (\gls{pcie}
Gen3 x8 vs. x16) does not appear to be a factor in these experiments.

\paragraph{\gls{pio} performance over \gls{pcie}}

In the second calibration experiment the CPU performs
\gls{pio} reads and writes over \gls{pcie} to the \gls{fpga}.
\gls{pcie} is complex, and we spent considerable time optimizing
CPU-initiated reads and writes for different platforms.  We describe
here how we arrived at optimally-efficient \gls{pio} latency to the
\gls{fpga} over \gls{pcie}.

We pre-map \gls{pcie} apertures (BARs) into user space on the CPU and
measure latency as the time to write a value to this space (thereby
\emph{invoking} a function on the device) and read a
result back (\autoref{listing:pciepio}) -- this is basically 
uncached \gls{mmio}.

\begin{figure}[t]
  \footnotesize
\begin{verbatim}
// map 4KiB of PCIe memory for read/write
int fd = open("/sys/bus/pci/devices/.../resource0",
    O_RDWR);
void *pcie_bar = mmap(NULL, 4096,
    PROT_READ | PROT_WRITE, MAP_SHARED, fd, 0);

// write arguments to PCIe memory
*(volatile invoke_args_t *)pcie_bar = invoke_args;
__sync_synchronize();

// read result from PCIe memory
invoke_res_t invoke_res = *(volatile invoke_res_t *)pcie_bar;
__sync_synchronize();
\end{verbatim}
\caption{Invoking with \gls{pio} \& \gls{pcie}; error
  handling omitted.}
\label{listing:pciepio}
\end{figure}

\gls{pcie} memory writes are \emph{posted} transactions and may complete
out-of-order, allowing multiple requests in-flight at the same time.
Most systems also have dedicated bus units to combine writes and coalesce
transactions.  Enzian's \txone performs \emph{write-combining} for
stores to \gls{pcie}, allowing 512 bits per bus round-trip
(confirmed with a logic analyzer) and resulting in efficient
\gls{pio} write transactions.

In comparison, \gls{pio} \emph{reads} over \gls{pcie} show low latency for small
data but degrade quickly as size increases.  This is because \gls{pcie}
memory reads are non-posted, forcing each read to finish
before the next can start and incurring a round-trip time cost (about
1\micros) for each word.  Many systems also have a narrow read
bus between CPU and \gls{pcie}; the \txone peripheral access bus
is only 128 bits wide.

This means that on \txone, common optimization techniques like vector
instructions have negligible benefit for \gls{pio} over \gls{pcie}:
the hardware already coalesces writes into 512-bit transactions and
reads are limited by the read bus width of 128 bits.

x86 has more optimization opportunities for \gls{pio} reads and writes.  The
Intel Data Streaming Accelerator~\cite{intel_intel_2024} offers
\texttt{MOVDIR64B} and \texttt{ENQCMD} instructions, allowing 64-byte \gls{pio}
writes with write-combining mappings and user-space descriptor
submission without driver intervention.  Mapping memory with
\emph{write-through} attributes (as in
Tide~\cite{humphries_tide_2024}), allows cache line-sized \gls{pio}
reads. These architecture-specific extensions are not available on
\txone. 

\autoref{fig:ev:together_pci} shows the best figures we could obtain
on both platforms.  For transaction
sizes above 32 bytes, the PC is about twice as fast as Enzian over the
\gls{pcie} cable between CPU and \gls{fpga}.
Here, the difference in \gls{pcie} bandwidth is responsible for the
latency difference, in particular for reads (limited to 128 bits on
both platforms).  Writes, in contrast, are combined and pipelined at
the \gls{pcie} interface.

Note this benchmark permits reordering of writes, and therefore
already reflects the performance achievable using the reorder unit
proposed in ~\cite{scenic-route}.

	\section{PIO over a coherent interconnect}\label{sec:opportunities}

While \gls{pio} over \gls{pcie} outperforms \gls{dma} on our
experimental hardware over a range of payload sizes (consistent with
the findings of other authors), \textbf{the main contribution of this paper to
show the further benefits of using a cache-coherent interconnect}.  We
present a family of message protocols, all of which run over an
existing MESI-like coherence protocol, which substantially outperform
conventional \gls{pio} and \gls{dma}.

\begin{figure}
  \includegraphics[scale=0.4]{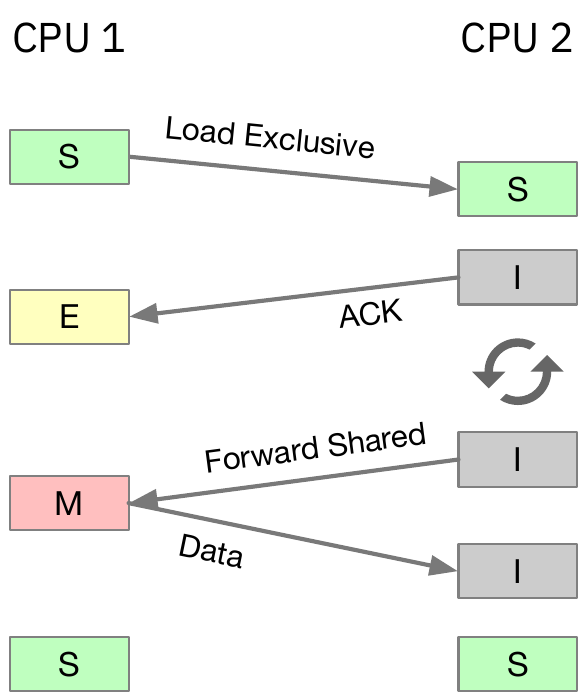}
  \caption{FastForward-style coherent messaging}
  \label{fig:ump}
\end{figure}

\paragraph{Fast CPU--CPU message passing}  Our starting point is a set
of software protocols used to communicate over shared memory between
CPUs in a cache-coherent system, including
FastForward~\cite{Giacomoni:FastForward:2008} and the \gls{nic}
emulation used in CC-NIC~\cite{cc-nic}, and exemplified in
\autoref{fig:ump}.  These target machines with
directory-based coherence which supports direct cache-to-cache
forwarding of lines without writebacks, e.g.\
MOESI~\cite{nagarajan_primer_2020}.

In these protocols, a receiving core generally spins reading from a
cache line (usually part of a larger array organized as a ring
buffer), holding it in cache in \Shared state.   The sending core
starts to write to the line, causing it to be fetched in \Exclusive
state and invalidated in the receiver's cache.  When the receiving
core next polls the cache line, it is fetched from the sending core
and arrives back in \Shared state on the receiver, where it is
immediately loaded into CPU registers.

In practice, the sender generally writes the entire line at once
(exploiting a write buffer), so that in the fast case the line is
transferred in two round-trips over the interconnect, as shown
in~\autoref{fig:ump}.  Techniques like including a ``finished'' flag
in the cache line data can be used to handle cases where the line is
transferred to the receiver before it has been completely written.

These protocols have excellent performance, and on modern hardware achieve
register-to-register latencies in the low hundreds of nanoseconds. They are
easily parallelized to many cache lines (with a single polled flag for many
lines of payload, plus the necessary barriers) to achieve excellent
throughput.  This basic technique is used, for example,
CC-NIC~\cite{cc-nic} to emulate a \gls{nic} using a different CPU socket.
However, they have several limitations.

The first is that the CPUs must busy-wait polling memory.  This does
not generate interconnect traffic since the line being polled is in
local cache, but does consume energy.  Moreover, optimal performance
depends on timing, and if the receiver polls too frequently, the line
can be forwarded multiple times before it is complete, leading to
wasted interconnect round-trips and increased latency.
For these reasons, these protocols are rarely used for real-world \gls{ipc},
even when it makes sense to dedicate cores to individual clients and
servers for long periods.

\paragraph{The implications of message-level access}

The key insight behind this paper is that a \emph{device} on a
symmetric coherent interconnect sees more cache state and
events than are visible to software, and can also initiate
more operations than a CPU core can.  Moreover, this does not require
a cache on the device at all, indeed this is rarely appropriate: it is
sufficient to generate and respond to individual coherence messages.
We explain this in more detail.

First, the device now \textbf{receives messages} from the CPU
cache that request lines in \Shared or \Exclusive state, or which request that
the caching state on the device be downgraded (e.g., from \Exclusive to
\Shared, or from \Shared to \Invalid).  It is free to react internally to
these events as it chooses, as long as it does not violate the protocol.

Second, the device can also \textbf{issue such requests} itself at any time,
allowing more fine-grained control of the state of a line in the CPU's cache.

Third, for lines which are homed at the device, the device' directory
maintains explicit \textbf{information about the cache line state} at the CPU
and other nodes as well as locally.

Fourth, \emph{unlike a conventional cache}, the device \textbf{does
  not have to respond immediately} to every request from the CPU's
cache.  Instead, it can choose to delay the response (blocking the
requesting core) until some other operations complete.  Care must be
taken to respond before any hardware-imposed timeouts, but these are
typically generous (hundreds of milliseconds on the \txone, for
example) and can be worked around in software (e.g., by sending a
``try again'' response within the timeout).

Finally, the device can \textbf{choose to interpret events} like remote
requests as more than simply loads or stores to the line concerned.  This last
insight is important because it, in combination with the previous observation,
it allows us to construct efficient communication mechanisms which rely on a
combination of the device's richer view of the protocol and the software on
the CPU adhering to additional conventions to communicate channel semantics to
the device.

This means requests to specific cache lines can be used to signal
particular higher-level operations to the device, much as register
access to traditional devices frequently has very different semantics
to memory access.

\newlength{\subfigwidth}
\setlength{\subfigwidth}{0.28\textwidth}
\begin{figure*}[t]
	\centering
	\begin{subfigure}[t]{\subfigwidth}
		\centering
		\includegraphics[width=\subfigwidth]{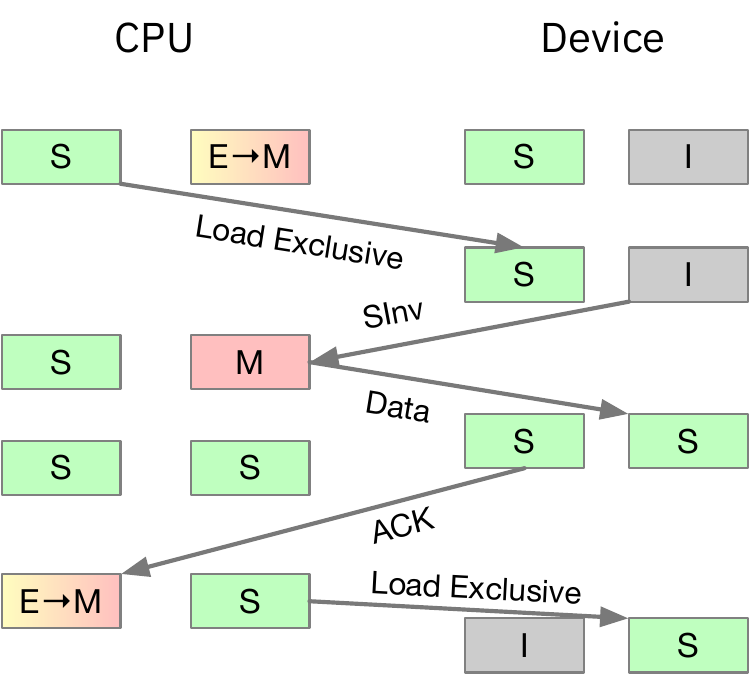}

		\caption{Write to device} \label{fig:protocols:cpu2dev}

	\end{subfigure}%
	\hspace{3em}%
	\begin{subfigure}[t]{\subfigwidth}
		\centering
		\includegraphics[width=\subfigwidth]{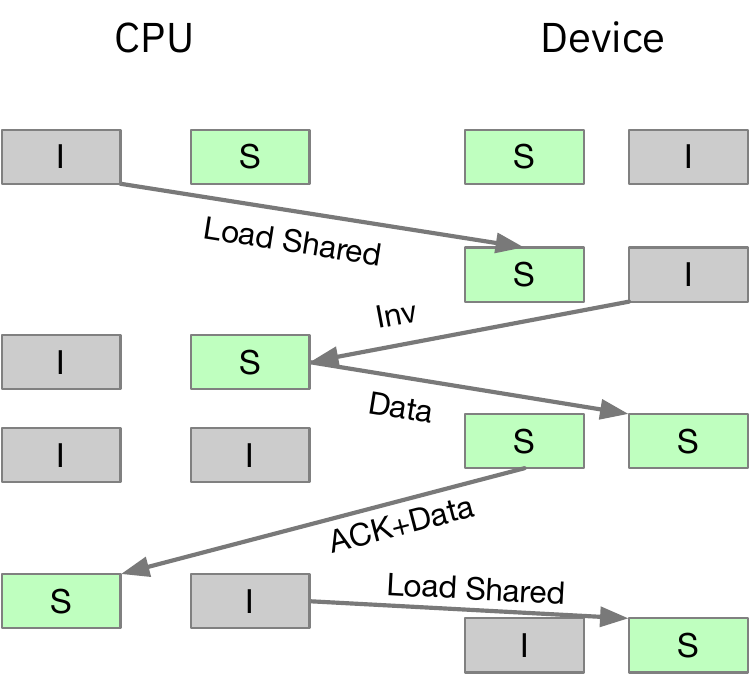}

		\caption{Read from device} \label{fig:protocols:dev2cpu}

	\end{subfigure}%
	\hspace{3em}%
	\begin{subfigure}[t]{\subfigwidth}
		\centering
		\includegraphics[width=\subfigwidth]{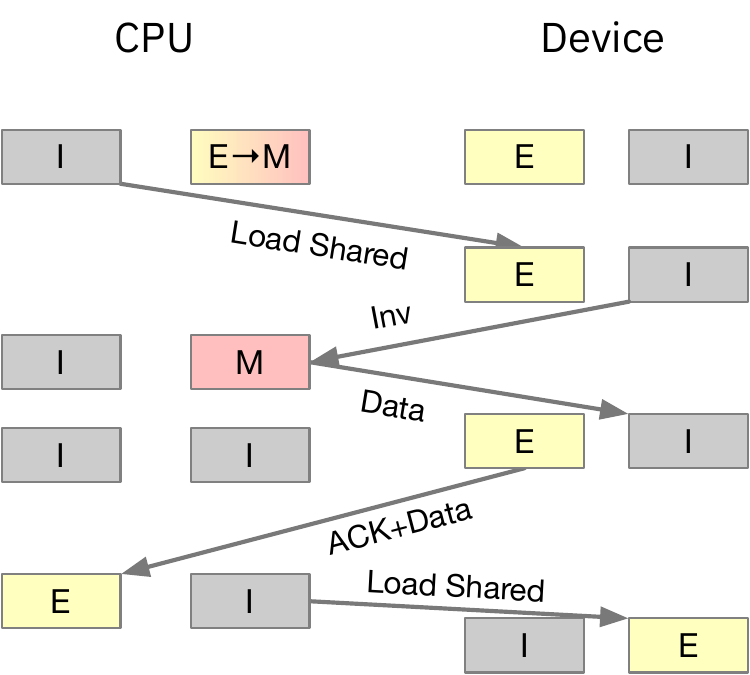}
		\caption{Read and write combined} \label{fig:protocols:bidirectional}
	\end{subfigure}

	\caption{Protocol variants for efficient CPU-device messaging} \label{fig:protocols}

\end{figure*}

\paragraph{CPU-device message passing with coherence}
We can now present a family of related point-to-point messaging
protocols which can be built over a symmetric, directory-based cache
coherence protocol where one end-point (the device) has direct access
to the low-level cache messages.

\autoref{fig:protocols} shows three such protocols, all with the same broad structure,
operating on a pair of cache lines which are homed at the device.
This allows the device to precisely track states of the cache lines
used and greatly simplifies the protocol.
They vary according to whether data flows (a) from CPU to device, (b)
from device to CPU, or (c) both ways.

Time is on the vertical axis.  Under each end-point (CPU or device)
are two columns, one for each cache line, giving its state over
time. The notation \texttt{E->M} within a cell indicates the silent
upgrade of a line from \Exclusive to \Modified, while the arrows
between cells indicate a state changes corresponding to visible
interconnect messages.

The first and second protocols  implement \emph{uni-directional}
message passing between the CPU and device, and are the basis for the
Ethernet \gls{nic} presented in \autoref{sec:eval:nic}.  The third
option combines the two by packing a request and response into one invocation of the
protocol (i.e.\ 2 round-trips) and is best suited for send-and-receive
interactions like \gls{rpc}.  We demonstrate this in \autoref{sec:eval:invocation}.

Each protocol requires just two interconnect round trips, beginning
with the two lines in a defined quiescent state. At each point one of
the lines (which we refer to as $A$) serves to transfer information
from CPU to device, and the other ($B$) from device to CPU. Over the
course of one transaction both lines migrate from one cache to the
other, ending in the same (caching protocol) state in which the other
began. The two then swap roles for the next iteration.

There is no actual cache on the device: any payload is delivered directly into
the device's execution pipeline, or read from an output register, as
appropriate. The following describes the execution of variant (c) in detail:
\begin{enumerate}
\item
    $B$ starts \Exclusive in the CPU cache, and $A$ \Invalid, with the
    (directory) states the opposite on the device.  Software writes payload
    data from registers into $B$, which (silently) transitions to \Modified.
\item
    To initiate the exchange, software reads from $A$, triggering a \emph{load
    shared} message to the device. The device interprets this as a signal that
    $B$ contains fresh data.

    This coupling of independent line states is the first deviation from
    `standard' coherence, and is impossible to achieve purely in software. It
    is nevertheless entirely invisible to the CPU.
\item
    The device now stalls the CPU by not responding to the load of $A$ until it has
    data to send. Instead, it requests $B$ in \Exclusive to fetch the payload,
    leaving it \Invalid in the CPU's cache.
\item
    The device uses the data in $B$ to computes a result which it returns as
    a result payload when it finally responds to the CPU's load from $A$.
\item
    The CPU cache receives $A$ and the CPU unblocks, immediately loading the
    first word of the response into a register.
\end{enumerate}

This \emph{would} lead to $A$ in \Shared in both nodes, since
software requested $A$ using a load.  However, as an
optimization, the device can return the line in \Exclusive instead,
invalidating its own copy, so the protocol returns to a
quiescent state with the roles of the lines reversed.

This protocol either eliminates or minimizes the drawbacks of a
software-only protocol: the ability to stall the CPU eliminates both
the need to poll, and the race condition. By coupling and overlapping
two coherence transactions we are able to transfer data in both
directions with the same two round-trips. All data
is transferred atomically so there is no risk of half a payload being
shipped.  This is the best achievable result with an unmodified
coherence protocol, since no cache transaction both sends and
receives a payload.

\begin{figure}[t]
	\includegraphics[]{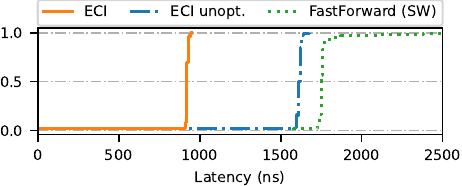}
	\caption{Invocation latency over \gls{eci} vs. FastForward}
        \label{fig:ev:rpc-cdfs}
\end{figure}

\autoref{fig:ev:rpc-cdfs} shows the distribution of total latency for this
operation implemented on Enzian.  Median latency is around
1600ns when the result is returned in \Shared state as requested by
the CPU (``ECI unopt'').  When we return the line in \Exclusive, this drops to
around 900ns (``ECI'').

For comparison, we also show an implementation of the FastForward
protocol~\cite{Giacomoni:FastForward:2008} exchanging cache lines between CPU
sockets in a dual-socket \txone-based Gigabyte R150-T61 server with similar
CPU and DDR specification to Enzian; this achieves a median latency of about
1750ns.

The optimized protocol is performing 2 round-trip message exchanges
over \gls{eci}, which has a one-way latency at the link layer of about
150ns.  The rest of the overhead (300ns) is mostly incurred in the
protocol processing in the directory controller, and would be
lower in an ASIC implementation.  The \gls{fpga} is clocked
at about 300MHz.

\paragraph{Returning a line in \Exclusive} The ability for the device
to unilaterally return a line in \Exclusive delivers an important
performance benefit.  The \gls{eci} protocol, and the \txone
last-level cache, both support this.  Interestingly, it
is also proposed in the (as yet unimplemented) CXL.mem
3.0 standard~\cite{cxl3cache}, along with the ``back invalidate''
operation providing symmetric cache coherence.  We suggest this is a
useful feature for future protocols.

\paragraph{Handling larger messages}

The protocols above handle one line at a time (128B on Enzian).  They
can be used repeatedly to transfer more data than this, but this is
inefficient.  Instead, we extend the protocols described so far using
parallelism and pipelining to give much better throughput.  This leads
to two further protocols, which are the ones we evaluate in
\autoref{sec:evaluation}.

For the \gls{nic} example of \autoref{sec:eval:nic}, with predominantly
unidirectional communication, we dedicate one pair of cache lines as
\emph{control} (as in \autoref{fig:protocols:cpu2dev} and
\autoref{fig:protocols:dev2cpu}) and add \emph{overflow cache lines} to carry
data.  When the protocol invalidates the control lines, it also invalidates these
overflow cache lines \emph{in parallel}, exploiting the full
bandwidth of the coherent link. We use this technique to transfer packets
larger than one cache line, up to an Ethernet jumbo frame (9kB).

We can do even better when the application has a symmetrical communication
pattern, i.e.\ the CPU sends and receives similar amounts of data in one
invocation.  This is the case for the device invocation and Timely Dataflow
experiments we show in \autoref{sec:eval:invocation} and
\autoref{sec:eval:timely-worked}. In this case, the better option is to
prefetch multiple instances of the protocol shown in
\autoref{fig:protocols:bidirectional}.

To transfer $n$ cache lines at a time, we maintain two groups of $n$
lines.  The CPU first writes all arguments for the device to
the lines in group 0, and then \emph{prefetches} all cache lines in
group 1. The prefetches trigger the device logic to invalidate all
cache lines in group 0, fetching the arguments.  The prefetches are
issued in parallel by the CPU and again provide enough concurrency to
saturate the link.

\paragraph{Handling timeouts} The protocols block a core's request
for a cache line until a device operation has completed.  This timeout
is typically quite long (often hundreds of milliseconds) but if it
happens the CPU's cache is likely to cause an exception, or other
fault like a machine check.

We solve this issue with a small state machine on the device that
returns a ``not ready yet'' response before an impending
timeout, causing the software on the issue another request for the
other cache line, extending the response time indefinitely
without the need for spinning.  In practice, we target operations that
are shorter than the timeout.

\paragraph{Avoiding deadlocks} It is also important to prevent
deadlocks due to the device stalling a request until it can issue, and
get a reply back for, another request to a different cache line.
Existing standards for inter-operable cache coherent interconnects are
silent on what happens in this situation, possibly because it has not
occurred until now. The design assumes that transactions on different
cache lines can progress independently.

In some implementations, this assumption holds in order
to maximize the memory bandwidth utilization and minimize request
latency, and also to simplify reasoning about deadlock freedom of the
coherence protocol itself.  However, it may be the case that the cache
stripes transactions across a limited number of independent units
which might deadlock if both $A$ and $B$ were mapped to the same unit.
Were this to be the case, it could be avoided by careful placement of
$A$ and $B$ in the physical address space.

This is actually the case in the Enzian \gls{fpga} directory
controller implementation we use~\cite{ramdas_cckit_2023}, which
follows similar provisioning choices in the CPU.
The \txone divides its last-level cache functionality into a set of
units called TADs, each of which can handle up to 16 simultaneous
cache transactions.  We ensure that consecutive cache lines are mapped
to different TADs to ensure that the operations can proceed independently.

Another possible source of deadlocks is that the CPU and L2 cache might issue
requests out of order, especially in the case of using prefetches.  This is
mainly relevant for handling requests larger than one cache line, as we
discussed above.  The device implementation needs to be aware of possible
reorderings and should not rely on the request ordering to perform state
transition to avoid deadlocks. In our implementation, we always buffer possible
in-flight cache lines on the \gls{fpga} and advance state machines based on
number of requests we see, without relying on the actual order of them.

\paragraph{Correctness and interoperability} An obvious concern given
the complexity of underlying cache coherence protocols is the
correctness of the implementation.  An error in the device might cause
the state machine in the CPU's cache to enter a state from which it
cannot recover.

Worse, while our implementation relies on an existing, and
well-tested, directory controller implementation on the \gls{fpga}, we
are using the \gls{eci} protocol in unconventional ways unforeseen by
the designers of the CPU and its cache.

In practice, after extensive testing we have not seen protocol-related
problems even under heavy load.  Moreover, we are reassured by recent
work on exhaustive semi-open-state testing of coherence
protocols~\cite{DBLP:conf/fmcad/SchultFCR24} which can formally define
the state space of a subset of a coherence protocol (which can
nevertheless include the kinds of behaviors we describe here), and
then exhaustively test a hardware implementation against that formal
model.

\paragraph{Enzian-specific implementation issues}
Our implementations will all be made available as open-source and use
a combination of SystemVerilog and SpinalHDL~\cite{spinalhdl}, and use
a version of the existing Enzian \gls{fpga} directory
controller~\cite{ramdas_cckit_2023}.


The ARMv8-A weak memory model means that barriers are needed to
reliably order reads and writes.  For example, it is critical that the
core's write buffer is drained into L1 cache (which is write-through
to the L2) before a subsequent load from the core signals to the \gls{fpga}
that the line can be pulled, and this requires a \texttt{DMB} barrier
instruction between the writes and the read.  On the \txone this is
sufficient to ensure ordering observed at the \gls{fpga}.

\paragraph{Generality} If we had only simulated the protocols in
\autoref{fig:protocols} rather than implement them in a hardware platform, it
seems unlikely that we could have become aware of the numerous
subtleties and challenges that real coherent interconnects present.
However, this raises generality concerns, since our
implementation is tied to a single platform.

In principle, all the protocols we have described here could be
implemented over any coherent interconnect which uses a symmetric,
MOESI-like, distributed directory-based model of coherence, and where
the interconnect hardware exposed major state transitions to the
device-specific logic which served as the end-point for messaging.

Several current and future standards satisfy these conditions, for
example CCIX~\cite{ccix}, TileLink~\cite{tilelink}, and CXL.mem 3.0~\cite{cxl3cache}, although
notably neither CXL 2.0 nor CXL.cache 3.0.  We are also seeing
something of a boom in custom interconnects from a range of vendors.
By making the requirements for this fine-grained, low-latency
communication clear, future designers can take this into account when
designing new interconnects and parts.

	\section{Evaluation}\label{sec:evaluation}

We compare coherent \gls{pio} with our two other
reference points: optimized \gls{pio} over \gls{pcie} described in the
previous section, and descriptor-based \gls{dma} over \gls{pcie}.
All use the same Enzian platform.  Coherent \gls{pio} uses the Enzian
native \gls{eci} implementation, while the \gls{pcie} approaches
use a loopback cable between \gls{pcie} interfaces on the CPU and
\gls{fpga}, effectively turning the Enzian \gls{fpga} into a
conventional \gls{pcie} \gls{fpga} card. \gls{dma} 
experiments use the Xilinx XDMA IP.

We explore three different application scenarios: synchronous
invocation to an accelerator on the \gls{fpga}, a closely-coupled high-speed
\gls{nic}, and a stream processor which moves some of its dataflow
operator graph to the \gls{fpga}.


With the exception of the \gls{nic} case, the results show very little
variance.  All graphs show 95\% confidence intervals, but these are
typically not visible. 

\subsection{Accelerator invocation}\label{sec:eval:invocation}

\begin{figure}[t]
	\includegraphics[width=\columnwidth]{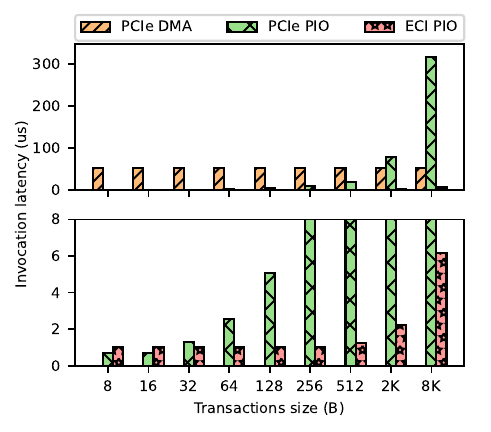}
	\caption{Invocation latency for different payload sizes}
        \label{fig:ev:together_thunderx}
\end{figure}

We first explore the scenario where the CPU invokes a function on the device configured as an
accelerator, using the three communication approaches to send argument
data and receive result data.   In this case, the \gls{fpga}
is configured to map the invocation into a write to, followed by a
read from, onboard Block RAM.

For \gls{dma} numbers we use the Xilinx XDMA
benchmark~\cite{xilinxbench}, which performs the operation using
descriptor-based \gls{dma}. We implement coherent \gls{pio} with the
protocol shown in \autoref{fig:protocols:bidirectional} to send a
request and receive a response in two round-trips.


\paragraph{Latency comparison} We first compare the latency for the CPU to
complete a function invocation on the device, measured as the time
taken for the CPU to send a request from the CPU to the \gls{fpga} and
read a response.  We vary the payload size (of both request and
response) and report median latency figures.

The results are shown in \autoref{fig:ev:together_thunderx}.  Up to
payloads of 8KiB, the latency of \gls{dma}-based communication is
dominated by descriptor setup and manipulation, and remains
largely independent of payload size (upper scale).  Faster \gls{pcie}
interconnects or CXL would only amplify this effect. 

Latency over \gls{eci} is dramatically lower across the board. It is
constant up to 256 bytes (two Enzian cache lines) and then increases
slowly, but linearly.  This is due to the pipelined nature of the
\gls{eci} \gls{pio} protocol: before we saturate the links, the
incremental latency induced by each extra line is small. 

Relative to \gls{eci}, \gls{pio} over \gls{pcie} performs poorly for
any payload more than 16 bytes, due to the inherently-limited
\gls{pio} read bandwidth discussed in \autoref{sec:platform}, but
competes well with \gls{dma} up to 2KiB payload.

We conclude that, for almost all transfers up to and beyond 8 KiB,
coherent \gls{pio} is significantly lower latency than both the
\gls{dma} option commonly used for these transaction sizes, and
\gls{pio} over \gls{pcie}.
More \gls{pcie} bandwidth (e.g. \gls{pcie} Gen5) will not change the
\gls{pio} results significantly without support for more efficient
\gls{mmio} reads, since \gls{pcie} round-trip latency remains the
same.  Neither would it improve \gls{dma} latency, which is dominated
by descriptor overhead.  In contrast, the lower interconnect latency
available with newer CXL versions \emph{would} improve things, but would also
deliver the same benefit to the coherent \gls{pio} case.

\begin{figure}[t]
	\includegraphics[width=\columnwidth]{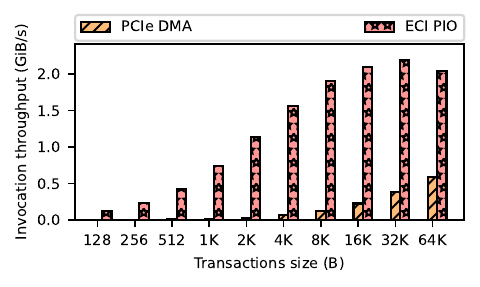}
	\caption{Invocation throughput for different payload sizes}
	\label{fig:ev:tput-comp}
\end{figure}

\paragraph{Throughput comparison} We also show throughput figures,
omitting \gls{pio} over \gls{pcie} since performance (as seen above)
is not practical for larger transfers.
We measure how many back-to-back invocations one CPU core can make for 1
ms, using different transaction sizes.  We calculate throughput by the total
data volume transferred divided by the time elapsed and report the median
value in \autoref{fig:ev:tput-comp}.

\gls{eci} \gls{pio} throughput increases with transaction
size up to 32KiB payloads to a peak of 2.19 GiB/s and then
dropping slightly, due to thrashing in the 32KiB L1 data
cache. 
It performs better than \gls{pcie} \gls{dma} over all
transaction sizes presented, and by a comfortable margin.  At 64 KiB,
we are still far from saturating throughput for \gls{pcie} \gls{dma};
it is well known that very large transaction sizes are needed to
achieve maximum throughput. 

In conclusion, \gls{eci} \gls{pio} shows a clear throughput advantage over
\gls{pcie} \gls{dma} at multiple page sizes (up to 64KB) for device
invocations, to the extent that the determining factor is less the
throughput achieved and more the number of CPU cycles or
cache size available.

As with the latency, this is unlikely to change with faster \gls{pcie}
interconnects, and CXL would likely result in improvements across the
board, including for the coherent \gls{pio}.

The results also show that there is an \emph{optimal} native transaction size
for coherent \gls{pio}, in this case the L1 cache size: larger transfers should
be broken down into smaller transactions of optimal size to achieve maximum
throughput.

\subsection{Network interface adapter} \label{sec:eval:nic}

\begin{figure*}[t]
    \centering
	\begin{subfigure}[t]{.3\linewidth}
		\centering
		\includegraphics[height=2.5cm]{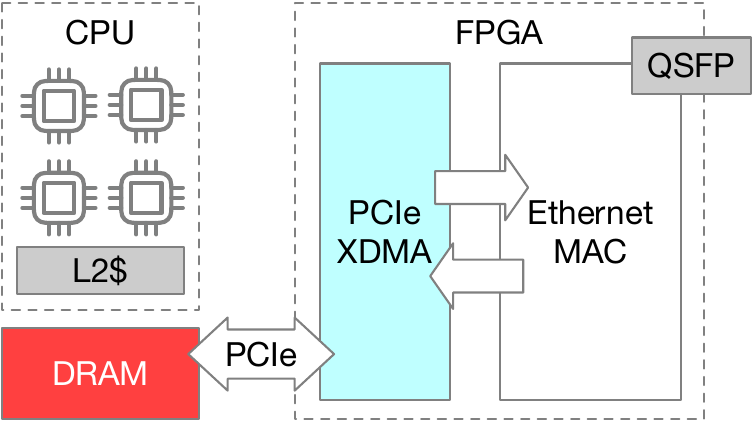}

		\caption{PCIe XDMA: packets transferred using
                  descriptor-based \gls{dma}}
                \label{fig:nic:xdma}

	\end{subfigure}%
	\hspace{1em}%
	\begin{subfigure}[t]{.3\linewidth}
		\centering
		\includegraphics[height=2.5cm]{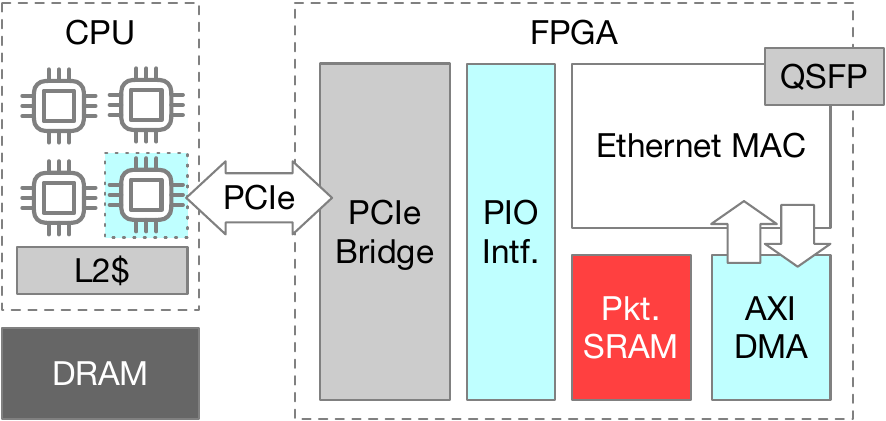}

		\caption{PCIe \gls{pio}: CPU reads descriptors and data from \gls{fpga} SRAM.}
                \label{fig:nic:pio}

	\end{subfigure}%
	\hspace{1em}%
	\begin{subfigure}[t]{.3\linewidth}
		\centering
		\includegraphics[height=2.5cm]{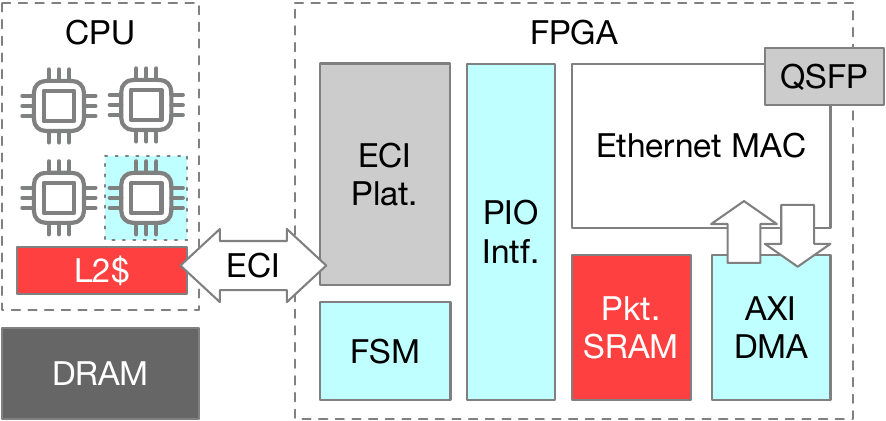}

		\caption{ECI \gls{pio}: CPU interacts with \gls{pio} interface and
		SRAM through the L2 cache.}
                \label{fig:nic:eci}

	\end{subfigure}%

	\caption{Different \gls{nic} architectures.  Blue denotes data movers; red
	denotes memory that holds packet data.} \label{fig:nic-archs}

\end{figure*}

We now compare approaches in the context of data center
networking.   Latencies of less than 1\micros are seen in data center
packet switching and
delivery~\cite{kachris_optical_2013,handley_re-architecting_2017,gibson_aquila_2022},
motivating minimizing latency between the network and 
CPU registers~\cite{pnet}.  We show that \gls{eci} \gls{pio} is a good
fit for such scenarios.

We implement 3 variations of a 100 Gb/s \gls{nic} in Enzian. The first
(\autoref{fig:nic:xdma}) is a conventional approach connecting one of Enzian's
100 Gb/s Ethernet MACs directly to the XDMA engines, resulting in a \gls{nic}
programmed using descriptor rings.  The second design
(\autoref{fig:nic:pio}) buffers packets in onboard SRAM on the \gls{fpga} and allows
the CPU to read and write packet and control data using \gls{pcie} reads and
writes.  The final approach (\autoref{fig:nic:eci}) uses a module on the
\gls{fpga} to bridge between the \gls{eci} directory controller and the MAC.
A variant of the protocol in \autoref{fig:protocols:dev2cpu} delivers network
packets to the CPU in multiple cache lines, and
\autoref{fig:protocols:cpu2dev} is used to send a packet. The 
\gls{nic} logic is clocked at 250 MHz. 

\begin{figure}[t]
    \centering
    \includegraphics[width=\linewidth]{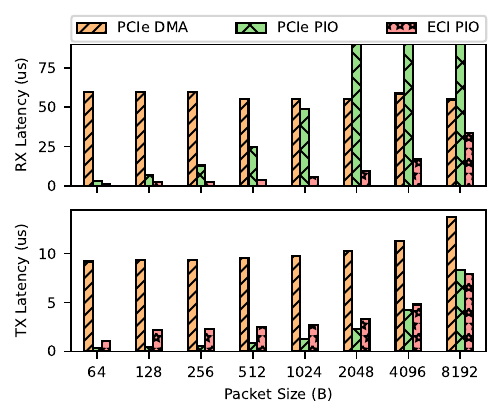}
    \caption{Latency of \gls{nic} implementations}
    \label{fig:ev:nic-rxtx-comparison}
\end{figure}

\newcommand{\mr}[2]{\multirow{#1}{*}{#2}}
\newcommand{\mc}[2]{\multicolumn{#1}{c}{#2}}

\newcommand{\bd}[1]{\textbf{#1}}

\begin{table}[t]
    \caption{\gls{nic} implementations: latency percentiles on selected packet sizes.
    Higher tail latencies are marked in \bd{bold}.} \label{tab:ev:nic-percentiles}
    \footnotesize
    \addtolength{\tabcolsep}{-1pt}
    \begin{tabular}{lrrrrrrrr}
        \toprule
        \mr{2}{Size} & \mc{4}{Percentiles RX (\micros)}              & \mc{4}{Percentiles TX (\micros)}             \\
                       \cmidrule(lr){2-5}                              \cmidrule(lr){6-9}
                     & P50     & P95     & P99         & P100        & P50     & P95     & P99         & P100       \\
        \midrule
        \cmidrule{1-9}
        \multicolumn{9}{l}{\textbf{PCIe \gls{dma}}} \\
                 64  & 65.39   & 66.36   & 67.65       & \bd{100.01} & 10.06   & 10.35   & 10.59       & \bd{16.49} \\
                 1536& 64.77   & 65.59   & 66.21       & \bd{133.84} & 10.89   & 11.12   & 11.33       & \bd{30.84} \\
                 9600& 65.89   & 67.12   & 68.05       & \bd{123.61} & 15.73   & 16.05   & 16.29       & \bd{41.99} \\
        \cmidrule{1-9}
        \multicolumn{9}{l}{\textbf{PCIe \gls{pio}}} \\
                 64  & 3.25    & 3.26    & 3.27        & 3.39        & 0.34    & 0.36    & 0.38        & \bd{4.80}  \\
                 1536& 72.89   & 72.93   & 72.96       & 73.05       & 1.82    & 1.84    & 1.86        & \bd{6.40}  \\
                 9600& 450.28  & 450.35  & 450.38      & 451.10      & 9.91    & 10.01   & 10.07       & 10.14      \\
        \cmidrule{1-9}
        \multicolumn{9}{l}{\textbf{ECI \gls{pio}}} \\
                 64  & 1.05    & 1.06    & 1.07        & 1.17        & 1.06    & 1.12    & 1.14        & 1.18       \\
                 1536& 7.24    & 7.29    & 7.39        & 7.43        & 3.09    & 3.26    & 3.50        & 3.59       \\
                 9600& 39.43   & 39.48   & 39.50       & 39.55       & 9.07    & 9.19    & 9.65        & 9.95       \\
        \bottomrule
    \end{tabular}
    \addtolength{\tabcolsep}{1pt}
\end{table}

For experiments we configure the Ethernet MAC in near-end PCS/PMA
loopback mode to reliably measure packet delivery
\emph{inside} the server without network delays.
We define \emph{receive latency} as time from the last beat of
the packet appearing on the ingress streaming interface of the
MAC, to the CPU fully receiving the packet contents in its
registers.  Transmit latency is similarly defined
as time from the CPU having the packet ready in
registers, to the last beat of the packet appearing on the egress
streaming interface on the MAC.  We do not include the time for
the packet to go through the Ethernet PCS/PMA loopback, since the MAC
is fixed and the same in all implementations presented.

To minimize disruptions from the Linux scheduler, we follow the common practice
to free one CPU core from kernel tasks and IRQ
processing~\cite{akkan_stepping_2012}, with the \texttt{isolcpus} function
together with core pinning using \texttt{taskset}.  In this setup, the only
disturbance from Linux would be the 250 Hz timer interrupt present on all
cores.  While a custom, completely \emph{tickless} kernel would remove this
disturbance completely, it is not commonly deployed in production environments
due to performance complications.  We use such a tickless kernel to
measure tail latency below, but other experiments use a stock
\texttt{generic} kernel from Ubuntu.
We show figures for XDMA in both polled and interrupt-driven modes; in
practice the difference in latency is small.

\autoref{fig:ev:nic-rxtx-comparison} shows receive latency for different
packet sizes for each \gls{nic} implementation.  As before,
\gls{dma} latency is stable across all packet 
sizes, suggesting that it is dominated by various \gls{dma}
overheads, such as descriptor setup and manipulation.
Note that system calls here actually result in little overhead
(only a few \micros); the overhead is dominated by cache
misses manipulating \gls{dma} data structures in RAM.

In contrast, \gls{pio} over both \gls{eci} and \gls{pcie} offer much lower
latency for small packet sizes up to 1024 bytes \gls{dma}, highlighting high
efficiency gains in \gls{pio} compared to \gls{dma}.  However, the latency of
\gls{pcie} \gls{pio} quickly degrades for larger packets due to the
\emph{non-posted read} problem discussed in \autoref{sec:platform}, reaching
over 350 \micros at 8 KiB packets.   In contrast, \gls{eci} offers
dramatically better latency even at this size.

\autoref{fig:ev:nic-rxtx-comparison} also shows transmit latency. 
\gls{pcie} \gls{pio} competes well with \gls{eci} across all packet sizes due
to combining posted writes.
\gls{pio} over \gls{eci} has slightly higher
latency for small packets, since it needs 2 round-trips over \gls{eci} for
each 128 bytes, versus a single \gls{pcie} round trip.  Both \gls{pio}
solutions perform significantly better than \gls{dma} across all packet sizes.

\autoref{tab:ev:nic-percentiles} shows the tail latency impact for selected
packet sizes for each \gls{nic} implementation.  We select three
representative packet sizes to evaluate tail latency: 64 for the smallest
possible packet, 1536 for the normal Ethernet MTU, and 9600 for a common jumbo
frame MTU.  We use the \emph{tickless} kernel setup mentioned earlier to avoid
the 250 Hz Linux timer, in order to reveal the actual tail latency impact from
our \gls{nic}.

\gls{pio} over \gls{pcie} results in low tail latency
compared to the descriptor-based approach, while \gls{eci} \emph{completely
eliminates} tail latency in this scenario.

These results show clear benefits of using \gls{pio} vs.
\gls{dma}, even for large packets, and especially when using a
coherent interconnect.  The poor receive performance of
\gls{pcie}-based \gls{pio} for receiving packets is due to the
unposted-read limitation of \gls{pcie}, and disappears for transmit.
This advantage should translate to future CXL3.0 and PCIe Gen6 \glspl{nic}.

\subsection{Offloading Timely Dataflow}\label{sec:eval:timely-worked}

Finally, we look at hardware offload of operators in a real-time
stream processing system: Timely
Dataflow~\cite{murray_naiad_2013,timely}. 
Timely schedules operators in a user-defined dataflow graph to
evaluate complex functions over on data streams, operating in
variable-sized batches. Operators are independent and may be
arbitrarily complex, thus benefiting from hardware offloading where
the performance improvement outweighs the overhead of shipping data to the accelerator.

Timely also illustrates the flexibility of the \gls{eci}
\gls{pio} protocol: our offloading implementation the protocol in
\autoref{sec:opportunities} \emph{synchronously} to exchange
statistics between \gls{fpga} and CPU before and after processing a
batch, and operators \emph{asynchronously} for processing data in the batch in parallel. 

We modify the latest Rust implementation of Timely~\cite{timely} to
include the ability to partition its dataflow graph between software
operators and those implemented on the \gls{fpga}, with automatic
insertion of communication channels between the CPU and \gls{fpga} to
transfer both data and progress tracking information across the
hardware/software boundary.  We perform two experiments: an extreme
synthetic benchmark with little actual computation, and real operators in
the form of offloaded Bloom filters.

\paragraph{Communication-only graph} The synthetic benchmark uses
31-element pipeline graph of trivial filter operators (filters). This
large operator graph maximizes communication overhead in the form of
Timely progress tracking data, while the simple operators mean the
\gls{fpga} has little performance advantage (if any) over the CPU,
mimicking cases where small batch sizes are need to ensure data
freshness, as in Differential
Dataflow~\cite{mcsherry_differential_2013}.


\begin{figure}[t]
	\includegraphics[width=\columnwidth]{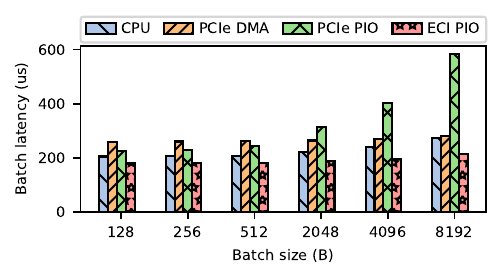}
        \caption{Timely synthetic filter graph offload}
	\label{fig:ev:cpu_xdma}
\end{figure}

The first (blue) bars of \autoref{fig:ev:cpu_xdma} show the baseline
performance of the non-offloaded (CPU-only) system against batch size. The
remaining bars cover the three offload cases.

\gls{eci} \gls{pio} batch latency is lower than both \gls{pio}
and \gls{dma} over \gls{pcie} for all batch sizes, even a single cache
line (128B).  Moreover, \gls{pio} over \gls{eci} is the only
technique that delivers lower latency than the software-only Rust
implementation, even in this near-worst-case scenario for hardware
offload. 

\paragraph{Bloom filters} In our second Timely experiment we offload
Bloom filters~\cite{bloom_spacetime_1970} to the \gls{fpga},
replicating the application scenario used to evaluate
Fleet~\cite{thomas_fleet_2020}.  Bloom filters are used to efficiently
check whether an element is a member of a set, using little memory.  A
Bloom filter consists of $k$ different hash functions and a lookup
table.  Testing the presence of an element requires computing these
$k$ hash functions and querying the lookup table with the results.

In the Fleet C++ software-only baseline, the hashes were computed in
parallel for each byte sequentially using AVX instructions.
We re-implemented this in Rust as a Timely operator using Arm SIMD
instructions. On the \gls{fpga} side, the hash computations are
parallelized and  iteration over the element's bytes is pipelined. 
We measure a single CPU thread and compare it with a single offloaded
operator. 

We implement the Bloom filter on the \gls{fpga} to take 128-byte
elements with $k=8$ hash functions that take byte-wide inputs.  We
unroll the hash function calculations for each byte lane by a factor
of 2, resulting in a 64-cycle latency for each element.  We then
pipeline the calculations with an initiation interval of 2 cycles to
saturate the 512-bit bus.  The return value is the set of 8 64-bit
hashes for each element.

In contrast to the synthetic benchmark above, this scenario has much
lower progress tracking overhead (only a single operator is offloaded,
compared with an entire sub-graph) and much higher computation cost
(the Bloom filters themselves).
The $k$ hash functions consist of sequential bit shifting,
addition, and XOR-ing, which are fast in \gls{fpga} logic and
deliver the offloading benefit even for small batches.

Instead of directly returning the hash results, the \gls{fpga} could
have computed the lookup table positions or performed the lookup
itself, returning either $k$ indices or binary values.  In the case of
indices, the read-to-write ratio would stay the same as in our
approach, but for the lookup, the read requests would be minimized and
writes would dominate. In this experiment, we assume that the lookup
table is located elsewhere and focus on efficiently calculating the
hash function and transferring the results to the lookup table.

\begin{figure}[t]
	\includegraphics[width=\linewidth]{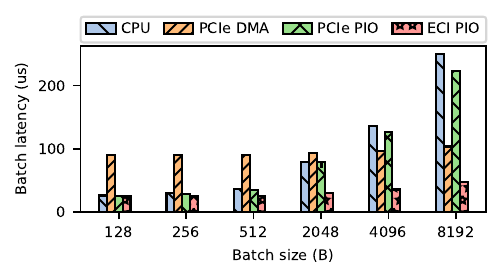}
	\caption{Offloading Bloom filters to the \gls{fpga}}
	\label{fig:ev:bloom}
\end{figure}

\autoref{fig:ev:bloom} shows the results.  Latency for both \gls{pio}
solutions and software (CPU) all start around 25 \micros, due to the
high overhead of streaming the input data. The per-element processing
time is 2.6 \micros on the CPU and 1.7 \micros when offloading with
\gls{eci}.  As the batch size (and hence processing time) increases,
the advantage of hardware offload increases, but so does the cost of
\gls{pio} over \gls{pcie}.  \gls{dma} latency barely
increases, but remains significantly higher than that for \gls{eci}.

As with the previous experiments, this shows the advantages of
the efficient, low-latency communication between CPU and device that
is possible using a cache-coherent interconnect, over and above both
\gls{pio} and \gls{dma} over an interconnect like \gls{pcie}.  We
anticipate that a future implementation of CXL.mem 3.0, which
similarly exploited cache coherence at the message level, would
deliver similar benefits.

	\section{Related work}\label{sec:related}

Cho \textit{et al.}~\cite{taming} provide an excellent survey of
existing CPU-device communication mechanisms, and address high-speed
storage devices with microsecond latencies.  They question the
conventional wisdom that \emph{hiding} the latency of interacting with
such devices is not possible using existing techniques, and show how
access latency using existing descriptor-based protocols can be
effectively hidden by clever use of prefetching, better hardware
queues, and user-level context switching.  Our work is different (and
complementary) in reducing the actual latency of each message between
CPU registers and device memory.

\subsection{Improving descriptor-based DMA}

Cohort~\cite{cohort} proposes a single, uniform, single-producer,
single-consumer, queue-based DMA descriptor interface to all
accelerators on an \gls{soc} which replaces the variety of ad-hoc
interfaces found on most \glspl{soc}.  Cohort mandates an MMU on each
accelerator, which maintains consistency with CPU MMUs and allows
seamless user-space access to accelerator queues, and integrates with
the cache coherence protocol to optimize the use of lock-free
descriptor queues.  We focus instead on what a latency-optimized
interface might look like, eschewing descriptors in favor of direct
cache line transfers.

Ensō~\cite{enso} shows the benefits of replacing the packet-based DMA
software interface used by modern \glspl{nic} with one based on a contiguous
large buffer of opaque, variable-sized messages, with metadata sent
via a different completion buffer.  Ensō replaces descriptors with
this buffer using an ingenious combination of direct \gls{mmio} writes
to the \gls{pcie} device and polling memory queues written by the \gls{nic}.
Our work similarly tries to reduce the overhead of descriptor
management, but by focusing solely on latency-sensitive small payloads
and avoiding DMA access completely.

\subsection{Coherent interconnects}

Coherent replacements or extensions of existing device interconnects
have been under development for some time and are beginning to see
widespread availability. IBM's OpenCAPI~\cite{OpenCAPI} builds on
\gls{pcie} by adding a protocol layer for coherence. It and the
competing Gen-Z~\cite{keeton_machine_2015} and CCIX~\cite{ccix}
protocols have meanwhile been either merged into or replaced by
CXL~\cite{Li:Pond:2023}, which seems likely to be the standard
interoperable protocol in the short- to medium-term future.  However,
commercial CXL 3.0 hardware implementing the more recent revisions has
yet to appear.

Other notable coherent interconnects include the RISC-V-specific
TileLink~\cite{tilelink,Terpestra:TileLink:2017,Cook:CARRV:2027} and NVIDIA's
NVLink 2.0~\cite{nvlink, nvlink1}. TileLink is an open standard and suitable
for research but hasn't yet been implemented in server-scale hardware. NVLink
is proprietary and closed.

Existing work has built on the available research platforms to explore
the design space for practical applications. Centaur~\cite{centaur}
demonstrated the \gls{fpga} offload of database operations on Intel
HARP v1.  However, while Intel HARP~\cite{harp1,harp2} coupled a CPU
and \gls{fpga} using a coherent QPI link, it provided a pre-configured
coherent cache in the \gls{fpga} rather than user access to coherence
protocol messages.  In contrast, this access is readily available in
Enzian~\cite{Cock:Enzian:2022}.

Open, extensible protocols have been used to explore coherent offload in
simulation, including making the case for protocol specialization building on
the Spandex protocol family~\cite{alsop2021case,spandex}. CoNDA~\cite{boroumand_conda_2019}
likewise employed simulation to explore the design space for coherent device
interconnects, with a focus on reducing unnecessary message traffic.

The Denovo protocol~\cite{denovo} is presented as an improvement specifically
for CPU-GPU coherence, tailoring the protocol to match the comparatively
predictable access patterns of typical GPU workloads.

\subsection{Cache line-based communication}

Communication in software between peer modes in a cache-coherent system is
typically very different from using descriptor-based DMA.
FastForward~\cite{Giacomoni:FastForward:2008}, Barrelfish~\cite{Baumann:Barrelfish:2009},
Concord~\cite{concord}, and Shinjuku~\cite{Shinjuku} adopt a much more direct
approach to sending small messages with low latency, e.g., by exploiting the
cache coherence protocol to transfer lines between caches on demand, and local
polling to provide synchronization.  Recently, cache coherence between NUMA
CPUs has been used to \emph{simulate} communicating between software and a
hypothetical cache-coherent \gls{nic}~\cite{cc-nic}.

Other work~\cite{ccix_pcie} has already noted that the cache line is a better
unit of transfer for small operations, as in FastForward~\cite{Giacomoni:FastForward:2008} and Barrelfish~\cite{Baumann:Barrelfish:2009} protocols. More recently
Concord~\cite{concord} communicates scheduling decisions between workers and
the dispatcher via a polled cache line, converting worker threads from
interrupt-driven to poll-mode ``CPU drivers''. A similar technique is used in
Shinjuku~\cite{Shinjuku}.

A recent study of data center \gls{rpc} from Google~\cite{google} reinforces the
importance of small transfers and highlights the large latency of \gls{pcie}
transactions.

HyperPlane~\cite{hyperplane} observes that I/O stacks in cloud
data centers frequently resort to spin polling the many descriptor
queues provided by modern \gls{nic} hardware, and proposes a a new
\texttt{QWAIT} hardware instruction which leverages the processor's
cache mechanism (much like \texttt{MWAIT}) to watch many different
in-memory descriptor queues at once, providing hardware acceleration
for \texttt{select()} or \texttt{epoll()}-like operations.  Evaluation
is performed using the Gem5 simulator, although this implementation is
not available.  We take a different approach based on transferring
data directly in cache lines without descriptors.

\subsection{Exploring PIO}

Neugebauer et.\ al.~\cite{pcie_perf} provide a comprehensive analysis of \gls{pcie}
performance in the context of \glspl{nic}. The space where PIO is preferable
to DMA for \gls{pcie} has been explored before. The
hXDP~\cite{slow-pci} \gls{fpga} \gls{nic} work highlighted the high overhead of
small \gls{pcie} transactions, and consequently performs small batch
computations solely on the CPU.

Larsen and Lee~\cite{larsen_reevaluation_2015} demonstrate the benefit
of write combining for \gls{pcie}-based PIO. Compared to a traditional DMA \gls{nic},
their \gls{fpga}-based prototype shows better latency and throughput for small-medium-sized messages and comparable throughput for large messages. The
\txone automatically performs combining for \gls{pcie} writes, and the results
of \autoref{sec:eval:invocation} thus reflect this optimization.  
Current work~\cite{scenic-route} proposes a write-reordering unit in the device to simplify
the use of fast \gls{pcie} posted writes and using Ensō~\cite{enso}
\gls{dma} to transfer data from the device. 

Dagger~\cite{dagger} builds on CCI-P, the commercialized implementation of
HARP's coherent cache, to construct an \gls{fpga}-based \gls{nic} specialized for low
latency \gls{rpc}.  A host-coherent cache holds connection states and the necessary
structures for the transport layer on the \gls{nic}, while the payload remains in
host memory. This minimizes \gls{fpga} memory demands, and exploits the lower
latency of cache misses compared to conventional \gls{pcie}-based \glspl{nic}..

	\section{Conclusion and Future Work}\label{sec:conclusion}

It is important to regularly question long-held assumptions about
systems, particularly in the light of new technological trade-offs and
emerging workloads.

We show that modern workloads can benefit from techniques that
optimize for small, frequent, low-latency interactions between CPU
cores and devices as much as the more traditional large,
throughput-oriented interactions for which current \gls{pcie} interconnects
and \gls{dma} are optimized.

Moreover, the use of cache coherent interconnects between CPU cores
and devices enables new techniques with this property, \emph{particularly}
when the device is able to participate in a standard cache coherence
protocol \emph{at the level of individual cache transitions} rather than
naive coherence.

We have demonstrated the potential of these techniques by showing a
family of messaging protocols implemented over Enzian's cache coherent
interconnect, but more importantly this opens up a wide design space
for the use of future coherent protocols like CXL.mem 3.0, which
offers many possibilities for more efficient fine-grained
communication within and between machines.

	\bibliographystyle{ACM-Reference-Format}
	\bibliography{references}

\end{document}